\documentclass[prb,floats,twocolumn,showpacs]{revtex4}
\usepackage{times,amsmath,graphicx,latexsym}
\begin{document}

\title{One-dimensional Excitations in Superfluid $^4$He and
$^3$He-$^4$He Mixture Films Adsorbed in Porous Materials}

\author{Hyung Cho}
\altaffiliation{Present address: Jet Propulsion Laboratory, 4800 Oak 
Grove Blvd., Pasadena , CA 91109.}
\author{Gary A. Williams}
\affiliation{Department of Physics and Astronomy, University of California,
Los Angeles, CA 90095}
\date{\today}

\begin{abstract}
A normal-fluid component varying as T$^{2}$ is observed at very low
temperatures in superfluid $^4$He and $^3$He-$^4$He mixture films
adsorbed in alumina powder.  The normal fluid appears to arise from
thermally excited third sound that has one-dimensional propagation
characteristics.  A Landau model of third sound excitations in an infinite 
cylindrical pore by Saam and Cole provides
good agreement with the experimental measurements over a wide range of
$^4$He and $^3$He coverages.  However, it is unclear why the powder 
substrate can be modeled as having cylindrical pores.

\end{abstract}

\maketitle

\section{introduction}

The Landau theory of superfluid $^4$He has been very 
successful in explaining the observed normal fluid density of the liquid
in terms of the thermal excitations, the phonons and 
rotons.\cite{tilley,khal}  At 
low temperatures only phonons are thermally excited, and a well-known 
result of the Landau theory is that the normal fluid density 
$\rho _n$ should then vary with temperature as\cite{khal} 
\begin{equation}
\rho _n={{2\,\pi ^2} \over {45}}\left( {{{k_B^4} \over {\hbar ^3\;c_1^5}}}
\right)\;\,T^4 \quad .
\label{eq:1}
\end{equation}

Here c$_{1}$ the velocity of first 
sound.  Experimental measurements at low temperatures are
consistent with this expression for the normal fluid 
density\cite{gw79, hanchu}, 
and an analogous T$^{3}$ 
dependence of the specific heat from the excited phonons is 
measured.\cite{phillips}

For two-dimensional (2D) helium films on a flat substrate, the comparable 
excitations are propagating thickness oscillations, known as third 
sound.\cite{atkins}  Density-functional theories\cite{dens} of thin films show 
that these excitations are well-defined and have linear dispersion
even for typical thermal wavelengths of order tens of \AA .
Applying the Landau model to this case gives an areal normal fluid 
density
\begin{equation}
\sigma _n={{2\,\pi ^2} \over {34.4}}\left( {{{k_B^3} \over {\hbar ^2\;c_3^4}}}
\right)\;\,T^3
\label{eq:2}
\end{equation}
where $c_{3}$ is the third-sound velocity.  This T$^{3}$ variation has 
been observed in measurements on helium films adsorbed on flat 
substrates.\cite{rutledge,saunders}

However, measurements of the normal-fluid density of helium films 
adsorbed in fine porous materials do not show the expected 2D behavior
as above.  This was observed in earlier measurements of films 
adsorbed in porous Vycor glass\cite{reppy} and in porous 
silica,\cite{takano} and is also seen in our measurements using an 
alumina powder substrate.\cite{cho,jltpa0,lt21,3he}  For these substrates the 
normal-fluid density is found to vary as T$^{2}$, indicative of a 
one-dimensional (1D) thermal excitation.  Computing the Landau theory for 
a 1D thickness oscillation yields for the 
normal-fluid density per unit length $\lambda _n$,   
\begin{equation}
\lambda _n={\pi  \over 3}\left( {{{k_B^2} \over {\hbar \;\,c_3^3}}} \right)\;T^2
\quad .
\label{eq:3}
\end{equation}
This is also the limiting result found by Saam and Cole \cite{saam} 
for third sound excitations in an infinitely long cylindrical pore.  
For a small pore diameter and very low temperatures the only modes 
excited are those propagating along the cylinder axis, since the 
modes transverse to the axis can only have wavelengths smaller than 
the pore diameter, and consequently high frequencies that make them 
energetically unfavorable at low T.

Since the torsion oscillator technique used in our experiments actually 
measures the superfluid mass per unit area, 
it is necessary to relate the linear and areal densities by assuming that the 
propagation is along a cylindrical pore of average diameter $D_{p}$, for which 
$\lambda _n=\sigma _n\,\pi \,D_{p}$.  The normalized areal superfluid density 
is then
\begin{equation}
{{\sigma _s(T)} \over {\sigma _s(0)}}=1-{{\sigma _n(T)} \over {\sigma _s(0)}}=
1-\beta '\,T^2
\label{eq:4}
\end{equation}
where the coefficient of the T$^{2}$ term is
\begin{equation}
\beta '={{k_B^2} \over {3\;\hbar \,\,D_{p}\;c_3^3\,\kern 1pt \sigma _s(0)}}
\quad .
\label{eq:5}
\end{equation}
In this paper we show that the Landau model for 1D excitations 
provides good agreement with our low-temperature measurements of the 
normal-fluid density in $^4$He and $^3$He-$^4$He mixture films 
adsorbed on a porous substrate.

\section{Experiment}
\begin{figure}[h]
\begin{center}
\includegraphics[width=0.45\textwidth]{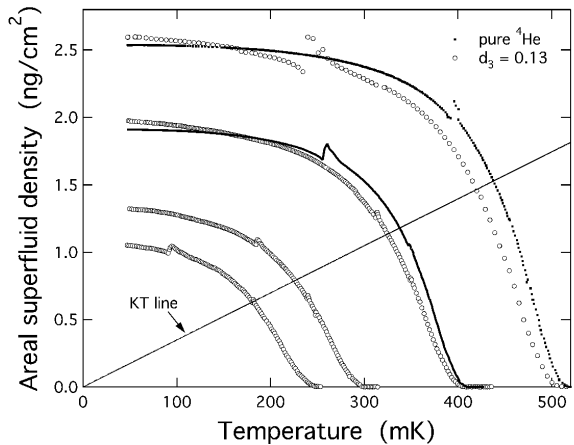}
\end{center}
\caption[]{Areal superfluid density versus temperature for two pure 
$^4$He films (solid dots) with $d_{4}$ = 0.50 and 0.38 layers (upper and lower 
curves), and four $^3$He-$^4$He mixture films (open circles) with $d_{3}$ = 0.13 
layers and (top to bottom) $d_{4}$ = 0.61, 0.49, 0.37, and 0.31 layers}
\label{fig1}
\end{figure} 
\begin{figure}[h]
\begin{center}
\includegraphics[width=0.45\textwidth]{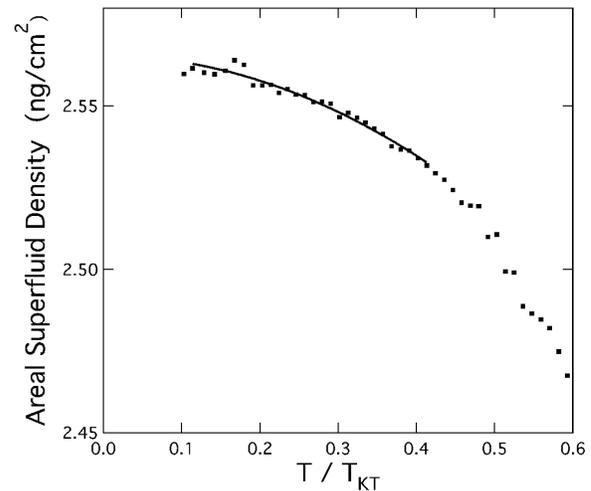}
\end{center}
\caption[]{Areal superfluid density at low temperatures for a film 
of pure $^4$He (from Ref.~\onlinecite{3he}) with a superfluid 
thickness of 0.55 layers and $T_{KT}$ = 453 mK.  
The solid line is the fit to Eq.~\eqref{eq:4}.}
\label{fig2}
\end{figure}
\begin{figure}[h]
\begin{center}
\includegraphics[width=0.45\textwidth]{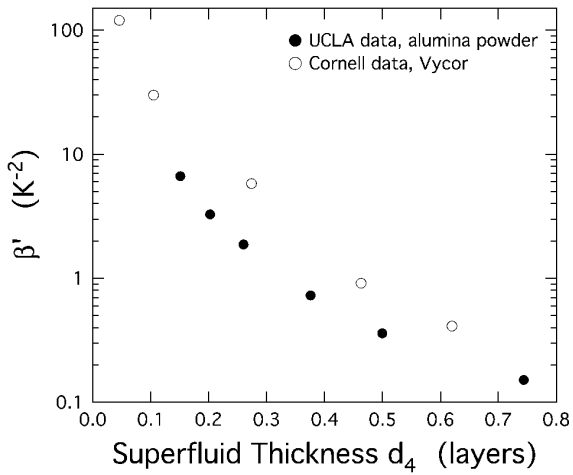}
\end{center}
\caption[]{The coefficient $\beta'$ of the $T^{2}$ term
as a function of the superfluid 
thickness for pure $^{4}$He films. 
The open circles are the porous Vycor glass data of Ref.~\onlinecite{mcqueeney}.}
\label{fig3}
\end{figure}
\begin{figure}[h]
\begin{center}
\includegraphics[width=0.45\textwidth]{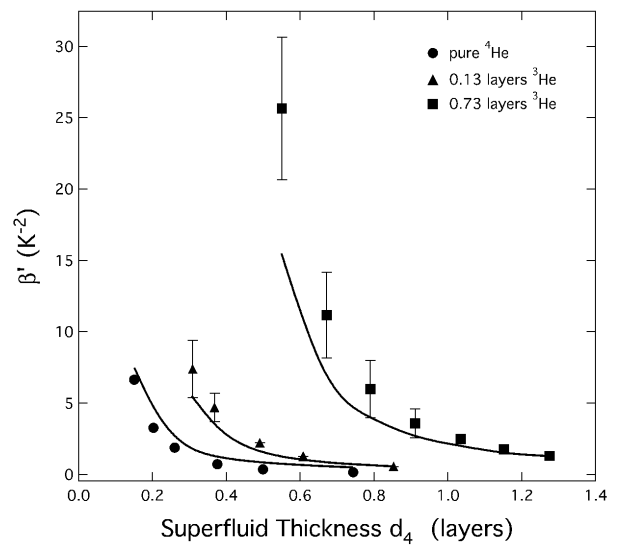}
\end{center}
\caption{$\beta'$ for $^{3}$He-$^{4}$He 
mixture films
as a function of the $^{4}$He superfluid coverage .  
The solid curves are the Landau theory of Eq.~\eqref{eq:5}.}
\label{fig4}
\end{figure}
\begin{figure}[h]
\begin{center}
\includegraphics[width=0.45\textwidth]{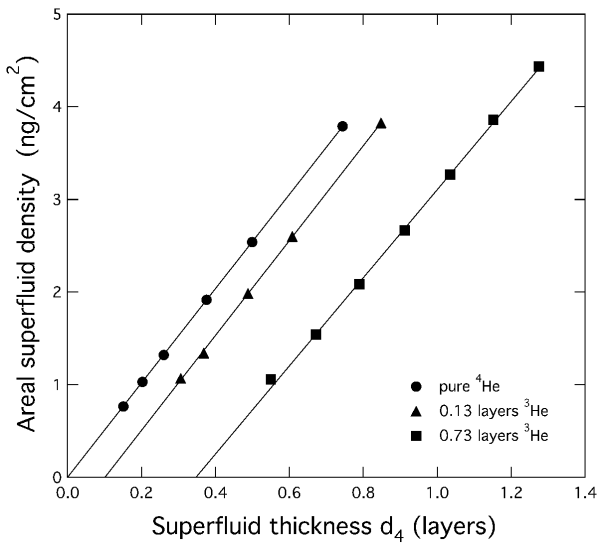}
\end{center}
\caption{Zero-temperature superfluid areal density as found from the 
fits of the data to Eq.~\eqref{eq:4}.}
\label{fig5}
\end{figure}
\begin{figure}[h]
\begin{center}
\includegraphics[width=0.45\textwidth]{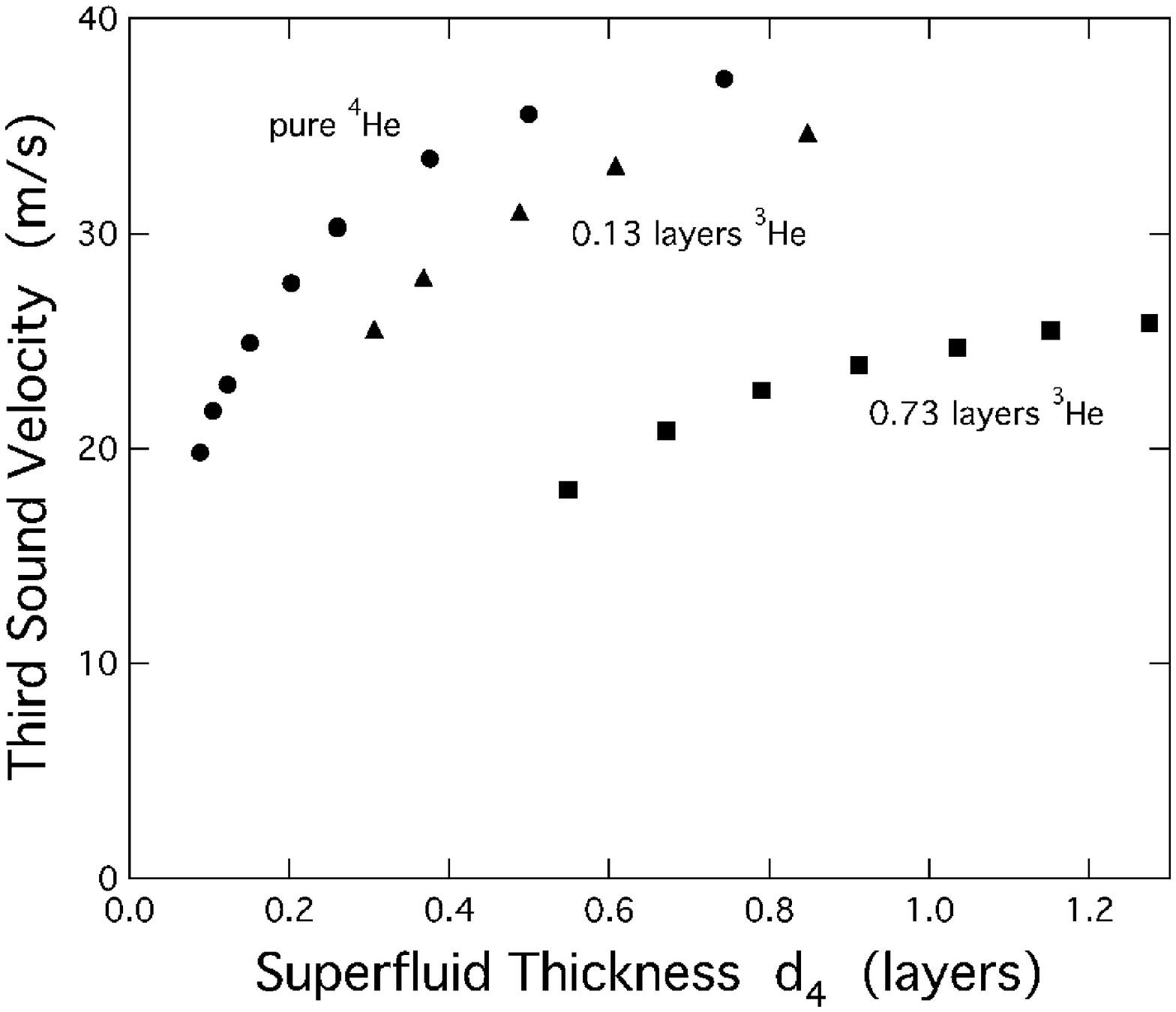}
\end{center}
\caption{Third sound velocities extrapolated to 
$T$ = 0.}
\label{fig6}
\end{figure}

The substrate used in our measurements is an Al$_{2}$O$_{3}$ powder of 
nominal diameter 500 \AA .  A slip-casting technique\cite{miguel}
uses the surface tension force of water draining out of the powder
to tightly pack the smaller powder grains into the voids around larger 
grains. This leads to a relatively low sample porosity (ratio of the open 
volume to the total volume) of P = 0.59.  A standard estimate\cite{jb}
of the pore size of our samples based on the porosity yields 105 \AA .
The volume of the sample is 5.07 cm$^{3}$, and its surface area
is 146 m$^{2}$.

The helium samples are condensed into the cell from measured amounts of gas
at room temperature, while the cell is maintained below 0.5 K with a 
dilution refrigerator.  The filling capillary of inside diameter 0.25 mm 
is 1 m long and is not thermally anchored except at 4.2 K and the 
sample chamber.  After condensing
$^3$He the sample is warmed to 4.2K for 24 hours to anneal the sample and 
assure uniform coverage of the $^3$He.\cite{chu} 
This step is not necessary when 
adding further superfluid $^4$He.  To 
characterize the helium coverages we take one layer of $^4$He to be
12.8 $\mu$moles/m$^2$ and one layer of $^3$He as 10.7 
$\mu$moles/m$^2$, corresponding
to the bulk liquid densities at low pressure.  
For the first few layers of the film this will be an overestimate of the 
actual number of atomic layers, since these are closer to the solid 
density due to the attractive substrate potential, but it should be a 
reasonable approximation for the top layer or two farthest from the 
substrate.  We find that the first 
2.7 layers of $^4$He are not superfluid at any temperature, forming the
inert "dead" layers known from earlier investigations.\cite{atkins}  We 
define a $^4$He thickness $d_{4}$ as the thickness in excess of the dead
layer; $d_{4}$ is then the thickness of the superfluid portion of the film at 
T = 0, with no $^{3}$He added.  A torsion oscillator is employed to 
determine 
the superfluid mass of the 
film, as described in Refs.~\onlinecite{cho} and~\onlinecite{3he}.  
By measuring the extrapolated $T$ = 0 period 
shift of the oscillator as a function of the mass of the helium added
the areal superfluid density of the films can be calibrated.  

\section{Results}

An example of our data for the areal superfluid density of both pure $^4$He 
and $^3$He-$^4$He mixture films is shown in Fig.~\ref{fig1}.  Further 
data can be found in References \onlinecite{cho,jltpa0,lt21,3he}. 
As shown by the correspondence with the 
critical Kosterlitz-Thouless (KT) universal line in this data, the 
superfluid transition of the films is mediated by thermally excited vortex 
excitations.  There is no sharp drop of the superfluid density to zero at
the temperature $T_{KT}$ where the data crosses the KT line because of the 
finite pore size of the substrate,\cite{pores}which acts to broaden the 
transition region.  The 
small glitches in the data of Fig.~\ref{fig1} are due to the 
third-sound modes of our cylindrical cell, which couple weakly to the 
torsional mode.  The glitches mark the mode-crossing points where the 
the third-sound frequencies match the torsional 
frequency as the superfluid density decreases with temperature.

Of interest in the present paper is the low-temperature regime of the 
data, $T$ $\leq$ 0.5 $T_{KT}$, where the density of thermally excited KT 
vortex pairs is negligible.  In this regime the data can be accurately 
fit to the form of Eq.~\eqref{eq:4}, and values of $\beta'$ can be extracted.
Figure~\ref{fig2} shows the low-temperature region for one data set 
of pure $^4$He from Ref.~\onlinecite{3he}, 
where the solid line shows the two-parameter fit with values 
of $\beta'$ = 0.36 K$^{-1}$ and $\sigma _s(0)$ = 2.565 ng/cm$^{2}$.
More general 
fits including terms linear and cubic in $T$ were tried, but the fit 
coefficients for such terms were negligibly small. 
Figure~\ref{fig3} shows the values of $\beta'$ obtained for a 
series of pure $^4$He films\cite{cho} of 
different thickness, plotted versus $d_{4}$.
Also plotted in Fig.~\ref{fig3} are the unpublished results of 
McQueeney\cite{mcqueeney} for a 
similar set of films in porous Vycor glass.  Both show a fairly rapid 
increase in $\beta'$ as $d_{4}$ (and $T_{c}$) is reduced, 
with the Vycor data being about 
a factor of two larger than the alumina powder results.  The higher 
values found in the Vycor sample are roughly consistent with 
Eq.~\eqref{eq:5}, which predicts that 
$\beta'$ should scale as the inverse of the average pore diameter: the 
pore diameter of the Vycor is thought to be about 75 \AA , while as noted 
above the average pore diameter of our slip-cast sample is of order 
100 \AA .  
The scaling with the inverse pore size also accounts for the considerably 
larger values of $\beta'$ seen with porous silica substrates\cite{takano} 
having 25 \AA \, pores, causing the $T^{2}$ term to dominate the superfluid 
density over nearly the entire temperature range up to the superfluid 
transition temperature.

In Fig.~\ref{fig4} our $\beta'$ data is plotted for $^{3}$He-$^{4}$He 
mixture films, where the $^{3}$He thickness $d_{3}$ is held constant and 
$d_{4}$ is varied by adding further $^4$He. The values of $\beta'$ are 
considerably increased with the addition of $^{3}$He, something which 
is readily evident comparing the mixture film curves in 
Fig.~\ref{fig1} with the curves for pure $^4$He at the lowest 
temperatures.  The error bars reflect the 
increasing uncertainties in the fits as $T_{c}$ is reduced with 
increasing $^{3}$He coverage, which puts an increasingly smaller fraction of 
the data at temperatures below 0.5 $T_{KT}$.  The largest error bar 
in Fig.~\ref{fig4}, for the film 
with $d_4$ = 0.55 and $d_3$ = 0.73, comes from the fits to the lowest 
curve in Fig.~\ref{fig1} of Ref.~\onlinecite{3he}.  With $T_{KT}$ = 139 mK for 
this data only a few points are available in the low-temperature 
regime for making a rough estimate of $\beta'$.

The fits to the data also give the T = 0 value of the areal superfluid
density, shown in Fig.~\ref{fig5} as a function of $d_{4}$.  As expected the 
variation is linear in the $^{4}$He coverage, but the addition of $^{3}$He
suppresses some fraction of the superfluid density.  Curves very 
similar to this have been observed for mixture films on flat Mylar 
substrates \cite{mylar}. 

To compare the data of Fig.~\ref{fig4} with the Landau model of 
Eq.~\eqref{eq:5}, it is necessary to measure the third-sound velocity.
This can also be obtained from 
the torsion-oscillator data, due to the third-sound glitches noted above.
These third-sound resonances are even more visible as 
dissipation peaks in the inverse Q factor of the oscillator, measured 
by monitoring the drive voltage needed to keep the amplitude of the 
oscillator a constant. Examples of this data are shown in Fig.~\ref{fig3} of
Ref.~\onlinecite{cho}.  For the cylindrical sample geometry with radius 
$R$ and length $L_{z}$, and taking the boundary 
condition that there be no heat flow from the outer surfaces of the 
powder sample, the resonant frequencies of the third sound are given 
by
\begin{equation}
\omega _{m\kern 1pt n\kern 1pt n_z}={{c_3} \over 2}\sqrt {\left( {{{\alpha 
^{\prime}_{mn}}
\over {\pi R}}} \right)^2+\left( {{{n_z} \over {L_z}}} \right)^2}
\end{equation}
where $\alpha ^{\prime}_{mn}$ is the 
n$_{th}$ zero of the derivative of the Bessel function J$_{m}$. The 
geometrical factors are known, and hence the third sound velocity can 
be determined from the oscillator frequency. Since the 
resonances are determined by the mode-crossing condition 
they occur at different temperatures for each mode, and in 
order to compare them the velocities are multiplied by the factor 
$\sqrt {\sigma _s(0)/\sigma _s(T)}$,
to extrapolate to the T = 0 value. It is found that the lowest 
five modes of a given film yield the same low-temperature speed 
to within about 5\% , and the average value is taken.  The
third sound speed c$_{3}$ is that for the film in the porous medium, 
and is smaller than that for the same film on a flat substrate 
by the index of refraction that accounts for the tortuosity 
of the multiply-connected film on the powder grains.\cite{heck}  
This can be deduced from the change in 
slope of the oscillator's period shift versus added helium when the 
film becomes superfluid; for the present slip-cast sample the index of 
refraction was 1.67.

Figure~\ref{fig6} shows the extrapolated third sound speeds at $T$ = 0 
versus superfluid film 
thickness for the pure $^{4}$He films and for two mixture film sets. The 
variation with superfluid thickness is just that expected for these 
very thin films where the Van der Waals restoring force dominates:  
a variation as $\sqrt {d_4}$
at small $d_{4}$, and then a turnover towards a maximum at 
larger thickness as the restoring force diminishes with increasing 
total film thickness.  
The addition of $^{3}$He lowers the third sound velocity since the
superfluid density is decreased,\cite{hallock} as in Fig.~\ref{fig5}. 

From the data shown in Figs.~\ref{fig5} and \ref{fig6} the predicted 
coefficient $\beta'$ of Eq.~\eqref{eq:5} can be 
evaluated for each of the film thicknesses shown in Fig.~\ref{fig4}; the solid 
lines in the figure are spline fits to the resulting values.
$D_{p}$ in Eq.~\eqref{eq:5} was adjusted to get the best match to the data, 
and the resulting value of 48 \AA \, is at least roughly consistent with our 
estimated pore size of 100 \AA.   As can be seen in Fig.~\ref{fig4} 
the agreement with the Landau 
expression for 1D excitations is quite good for the entire range of 
$^4$He and $^3$He coverages, considering that only the single parameter $D_{p}$ 
was adjusted to get the agreement.

\section{Discussion}

Although the Landau theory provides a good description of the 
experimental results, the reasons why the excitations display one-dimensional 
behavior in the geometry of 
the slip-cast powder remain unclear.  In the theory of Saam and Cole 
\cite{saam}
the cylindrical geometry with pore length much greater than the 
diameter is crucial to the appearance of 1D behavior at low 
temperatures.  There is no obvious reason why the pores in the powder 
can be modeled in this fashion.  The frequency of third sound with 
energy $k_{B} T$ at a temperature of 0.3K is $k_{B} T / 2 \pi \hbar$ = 
1x10$^{10}$ Hz, 
corresponding to a wavelength of order 20 \AA.  This is smaller than the 
pore diameter, and one might expect instead 2D propagation 
characteristics for that case, as seen on the flat substrates. 
An unknown factor is the mean free 
path of the high-frequency third sound, which could well be less than 
the pore size. 

One consideration which might be important is the orientation of the pores 
with respect to the direction of the superflow imposed by the 
torsional motion.  Quasi-cylindrical pores whose axis is parallel to the 
motion contribute considerably more to the measured superfluid 
fraction than those 
perpendicular to the motion; this is the origin of the index of 
refraction for the porous materials.  If the third-sound scattering 
process which transfers momentum to the substrate is also anisotropic
in the cylindrical channel, such that only propagation directions 
along the axis contribute, this could give possibly lead to a 
1D behavior of the superfluid density.

A further possibility is that the points where the grains make contact 
with each other may play a role.  All of the flow between the grains 
is channeled into these regions, which are probably smaller in extent 
than the $\approx$ 20 \AA \, thermal third-sound wavelength.  They would 
effectively be 1D channels, and since the flow velocity is locally 
much higher than the average, these regions might account for a 
disproportionate share of the reduction of the superfluid density.
It is clear that further theoretical work modeling 
superflow and third-sound propagation in a porous multiply-connected 
geometry will be needed to understand the observed effects.

An additional experimental signature of 1D propagation would be a 
low-temperature heat capacity linear in $T$.  The Landau model yields 
for N cylindrical pores of length L a heat capacity
\begin{equation}
C=NL\left( {{{\pi k_B^2} \over {6\hbar c_3}}} \right)T
\quad ,
\end{equation}
which is also the low-temperature limit of the calculation of Saam 
and Cole.  However, experiments on superfluid films in Vycor over the 
the thin-film range of coverages discussed here show heat capacities 
varying more as $T^{2}$.  The heat capacity is a more complicated quantity
which also has considerable contributions from the "dead" layers, and 
this may be obscuring the contribution from the 1D third sound 
excitations.

\begin{acknowledgments}
This work was supported by the National Science Foundation, Division 
of Materials Research, under grants DMR 95-00653  and  DMR 97-31523.
\end{acknowledgments}


\begin{thebibliography}{35}
\bibitem{hanchu}Han-Ching Chu and one of the present authors (G. A. 
W.) have carried out further measurements on the same alumina sample
discussed in this paper, but where the pores were completely filled with 
$^{4}$He.  Fitting to the decreasing superfluid fraction between 100 
and 600 mK for this case yields a power-law exponent with temperature of 
3.7$\pm$0.2, consistent with the value of 4 expected for 3D 
behavior.
\end{thebibliography}
\end{document}